\documentclass[]{spie}  

\usepackage{amsmath,amsfonts,amssymb}
\usepackage{graphicx}
\usepackage[colorlinks=true, allcolors=blue]{hyperref}
\usepackage{xspace}
\usepackage[table]{xcolor}

\def\m87{M87$^*$\xspace}

\def\lsim{\mathrel{\raise.3ex\hbox{$<$\kern-.75em\lower1ex\hbox{$\sim$}}}}
\def\gsim{\mathrel{\raise.3ex\hbox{$>$\kern-.75em\lower1ex\hbox{$\sim$}}}}

\title{The Black Hole Explorer Cryocooling Instrument}

\author[1,2]{Hannah Rana}
\author[3,7,2]{Kazunori Akiyama}
\author[4]{Edgar Canavan}
\author[4]{Michael DiPirro}
\author[1]{Mark Freeman}
\author[2,5,6]{Peter Galison}
\author[1]{Paul Grimes}
\author[7,8]{Mareki Honma}
\author[1]{Janice Houston}
\author[1,2]{Michael Johnson}
\author[4]{Mark Kimball}
\author[9]{Daniel Marrone}
\author[1]{Edward Tong}

\affil[1]{Center for Astrophysics $|$ Harvard \& Smithsonian, 60 Garden Street, Cambridge, MA 02138,
USA}
\affil[2]{Black Hole Initiative at Harvard University, 20 Garden Street, Cambridge, MA 02138, US}
\affil[3]{Massachusetts Institute of Technology Haystack Observatory, Westford, MA 01886, USA}
\affil[4]{NASA Goddard Space Flight Center, Greenbelt, MD 20771, USA} 
\affil[5]{Department of History of Science, Harvard University, Cambridge, MA 02138, USA}
\affil[6]{Department of Physics, Harvard University, Cambridge, MA 02138, USA}
\affil[7]{Mizusawa VLBI Observatory, National Astronomical Observatory of Japan, Iwate 023-0861, Japan}
\affil[8]{Department of Astronomy, Graduate School of Science, The University of Tokyo, 7-3-1
Hongo, Bunkyo-ku, Tokyo 113-0033, Japan}
\affil[9]{Department of Astronomy and Steward Observatory, University of Arizona, 933 N Cherry Avenue, Tucson, AZ 85721, USA}

\authorinfo{Corresponding Author: Hannah Rana (\url{hannah.rana@cfa.harvard.edu})}

\begin{document} 
\maketitle

\begin{abstract}
The Black Hole Explorer is a space-based very-long baseline interferometry (VLBI) mission that will seek to perform precision black hole measurements, detect the photon ring around a black hole, explore the spacetime, spin, and mass properties of black holes, and attempt to experimentally validate predictions of General Relativity. These ambitious goals are achieved through the use of cryogenic receivers offering quantum-limited sensitivities across a wide frequency coverage. The dual-band receivers at 80-106\,GHz and 240-320\,GHz require 20\,K and 4.5\,K operating temperatures, respectively. To reach this, the planned cryocooling system will include two cold stages; a 20\,K stage which must lift a heat load of approximately 125\,mW and a 4.5\,K stage lifting 10\,mW of heat load. A survey of 4\,K cryocooler development for spaceflight is explored in order to baseline the cryocooling system design for BHEX and leverage existing technology in the space industry at high TRLs. Notable space missions of relevance include Planck, JEM/SMILES, Hitomi, XRISM, and the advancement of US cryocoolers in this temperature range thanks to the ACTDP/JWST. Integration of the cryocooler with the receivers and broader instrument requires careful consideration, as it influences the instrument operation and thermal challenges. The latter includes thermally linking the cold ends of each cooling stage whilst minimising heat losses and ensuring adequate passive cooling for the cryocooler warm end heat rejection. Moreover, the challenges and trade-offs in sizing the mass and reducing the power consumption are explored: varying modes of operation in conjunction with other key instrument subsystems, the receiver cold temperature requirements, which in turn influence the scientific objectives of the mission, and mitigating the mission critical risks of the system. Overall, this paper presents an overview of cooling needs, initial design considerations, a survey of 4\,K spaceflight cryocooler developments and current progress, and balancing scientific requirements of the instrument with the limitations of technical cryocooling capabilities, within the framework of a small-class (SMEX) space mission aiming to achieve breakthrough goals in experimental black hole physics.
\end{abstract}
 
\keywords{Black Holes, VLBI, Photon Ring, Cryocooler, Stirling, Joule-Thomson, Pulse Tube, Receiver, EHT}

\section{Introduction}
\label{sec:intro}  
Black holes play a pivotal role in understanding stellar evolution, galaxy formation and mergers, astrophysical jets, and the nature of spacetime \cite{BHEX_Johnson_2024}.  Through the Event Horizon Telescope (EHT), a network of telescopes conducting very long baseline interferometry (VLBI), it has become possible to produce resolved images of black holes \cite{EHT2019a,SgrAEHTCI}. By extending the baseline of the EHT into space, the Black Hole Explorer (BHEX) aims to achieve multiples of higher angular resolution in order to resolve, for the first time ever, the photon ring of a black hole, as well as study spacetime, directly measure black hole spin and other properties, and investigate jet accretion mechanisms. Such superior direct study and imaging of black holes in space will undoubtedly lead us into a new era of discoveries about black holes and our universe. 

The BHEX mission is targeting a NASA Small Explorer (SMEX) mission class and will operate a space-Earth hybrid observatory with the EHT. The instrument consists of a 3.5\,m antenna \cite{BHEX_Sridharan_2024}, dual-band receivers for simultaneous observations \cite{BHEX_Tong_2024}, 100\,Gbps laser data downlink \cite{BHEX_Wang_2024}, stable frequency reference to ensure coherence with ground operations \cite{BHEX_Marrone_2024}, digital backend \cite{BHEX_Srinivasan_2024}, and the cryocooling system, which is detailed in this paper.

\section{Cryocooling Requirements}
\label{sec:reqs}
The BHEX receiver front-end will observe simultaneously in two bands \cite{BHEX_Tong_2024}. In order to reach the sensitivity needed for the scientific goals, the receivers require cooling down to cryogenic temperature. Firstly, a 80-100\,GHz channel will employ cryogenic low noise amplifiers, enabling high-frequency observations, and will require an operating temperature of 20\,K. Secondly, channels at 240-320\,GHz consisting of SIS receivers require an operating temperature of 4.5\,K.

There has been an increasing need in the space-based astronomy community for compact, efficient, and closed-cycle cryocoolers that can reach $\sim$4\,K cold temperatures in space \cite{rana2023}. A few astrophysics missions have utilized liquid cryogens to achieve cooling to 4\,K, however, these pose a series of disadvantages for SMEX missions, namely the added mass to the instrument payload. Closed-cycle cryocoolers within the range of 4-6\,K have flown on missions including Planck, JWST, and XRISM \cite{Morgante_2009, jwst_miri, XRISM_cryo}. These operate with a combination of Stirling and Joule-Thomson (JT) cooling. BHEX aims to leverage existing cryogenic technology, where the majority of the key cryogenic components within these technologies, such as Stirling and JT coolers, boast established spaceflight heritage.

Imported vibration from the cryocooling system to the receivers needs to be kept below $~$5$\mu$m. Keeping within a NASA SMEX mission, the total instrument budgets are typically around 300\,kg and 700\,W for mass and power, respectively. Hence, the cryocooling system must be able to keep within approximately 1$/$3 of these budgets. Table \ref{tab:requirements} summarizes the receiver cryocooling requirements, where $T_c$ is the operating cold temperature and $Q_h$ is the heat load to be cooled.

\begin{table}[ht]
\caption{BHEX Receiver Cryocooling Requirements} 
\label{tab:requirements}
\begin{center}       
\begin{tabular}{|l|l|l|l|} 
\hline
\rowcolor{green!30} \textbf{Component} & \textbf{Frequency} & \textbf{$T_c$} & \textbf{$Q_h$} \\
\hline
\rowcolor{green!5} \rule[-1ex]{0pt}{3.5ex}  
High Band Receiver & 240-320\,GHz & 4.5\,K stage & 10\,mW  \\
\hline
\rowcolor{green!5} \rule[-1ex]{0pt}{3.5ex} 
Low Band Receiver & 80-100\,GHz & 20\,K stage & 125\,mW  \\
\hline
\end{tabular}
\end{center}
\end{table}

\section{Technology Survey and Spaceflight Heritage}\label{sec:techsurvey} 

\subsection{Stirling and JT Cryocoolers}
\label{sec:StJT}
\begin{figure} [ht]
   \begin{center}
   \begin{tabular}{c} 
   \includegraphics[height=8cm]{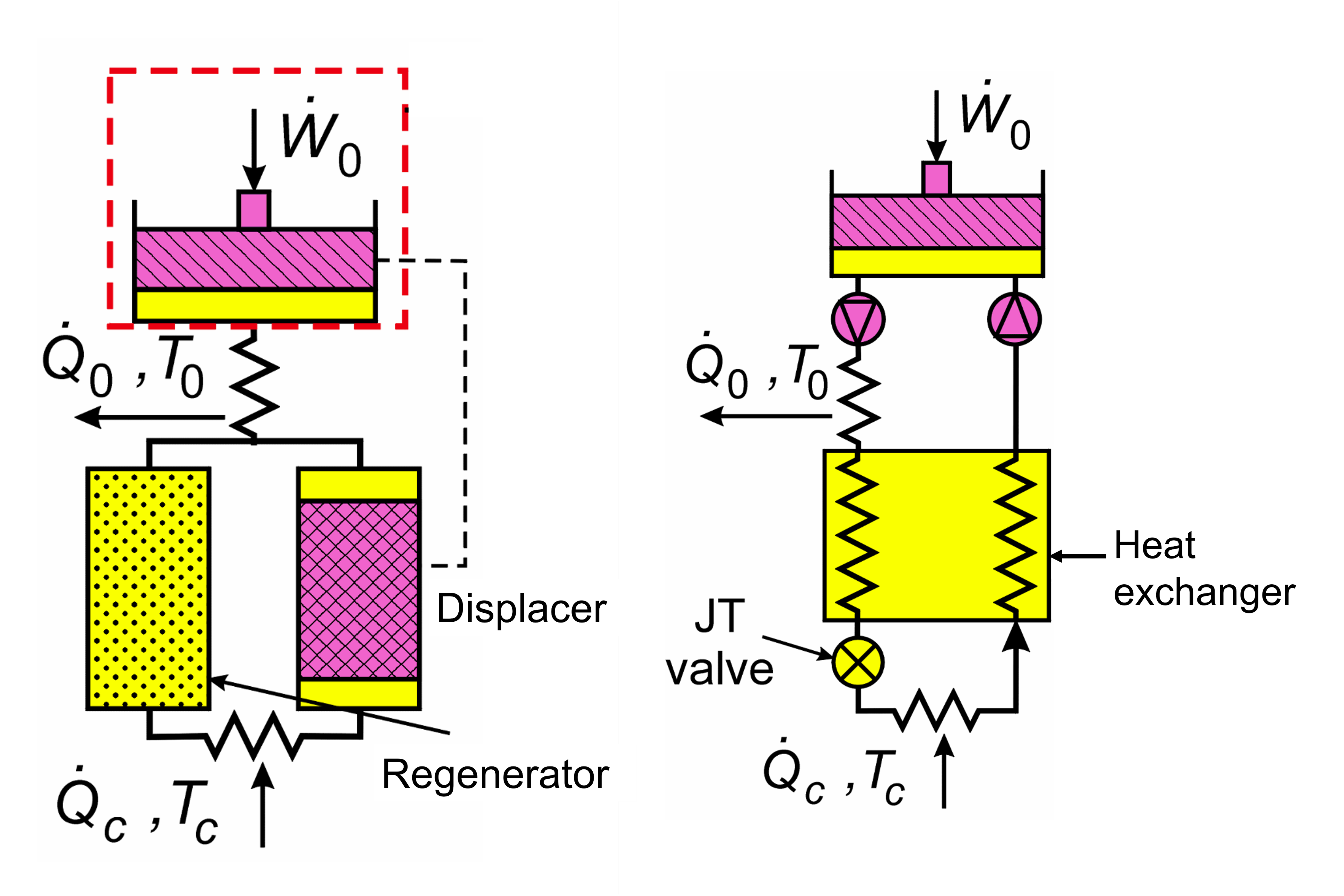}
	\end{tabular}
	\end{center}
   \caption[Schematics of the regenerative operation of the Stirling cryocooler and the recuperative cycle of the Joule-Thomson cryocooler (adapted from \citenum{radebaugh_2009}).] 
   { \label{fig:st-jt} 
Schematics of the regenerative cycle of Stirling cryocoolers and the recuperative cycle of Joule-Thomson cryocoolers (adapted from \citenum{radebaugh_2009}).}
   \end{figure}

Stirling and Joule-Thomson (JT) cryocoolers can be used together to achieve 4\,K cryocooling in space, where established designs utilize Stirling cryocoolers to reach an initial cold stage of pre-cooling and employ JT coolers to reach a lower 4\,K stage. Stirling cryocoolers boast being the most utilized spaceflight cryocoolers, yielding high efficiencies and long lifetime performance for space missions \cite{radebaugh_2009, rana2023}. In addition to superior flight heritage, recent research has achieved low cold end vibration for spaceflight Stirling pulse tube cryocoolers (SPTCs) designs \cite{cryoblue, passive1, passive2}. Figure \ref{fig:st-jt} presents schematics of the processes governing the Stirling cycle cryocooler and the JT cryocooler operation. 

The Stirling cryocooler runs a regenerative cycle and is driven by oscillating flows and pressures, and predominantly uses high-pressure helium as the working fluid. The operating frequencies range from 60-100\,Hz for Stirling drive compressors. In regenerative cryocoolers, heating occurs as the pressure increases, and cooling occurs as the pressure decreases. A displacer at the `warm end' moves most of the gas to the hot end during compression and to the cold end during expansion. The displacer is further responsible for controlling the phase difference between the compressor and displacer oscillation and maximizing the refrigeration power that can be achieved in the thermodynamic cycle. The JT cycle is categorized as a recuperative cycle, where gas flows steadily in one direction with consistent low and high pressures in specified locations. In this cycle, compression occurs at the ambient temperature ($T_0$), with the heat generated from compression being dissipated to the surroundings. Expansion takes place at the cold end at a temperature $T_c$, where the net refrigeration power $\dot{Q_c}$ is absorbed. The consistent flow and pressure in this cryocooler result in minimal temperature oscillation and vibration, especially when rotary compressors are used. The JT cycle, having no moving parts at the cold end, can be easily scaled to micro-sizes \cite{bradley_2009, microJT_book}. To reach cryogenic temperatures, high-efficiency heat exchangers are essential due to the relatively small adiabatic temperature changes associated with constant enthalpy expansion in the JT cycle compared to the temperature difference between ambient and cryogenic temperatures. Typically, heat exchanger efficiency must be at least 95\% to achieve net refrigeration at cryogenic temperatures when using high pressures \cite{dewaele_JT, radebaugh_2009}.

The majority of 4-10\,K cryocoolers developed for space applications use a combination of Stirling and JT coolers and will form a significant part of the discussion to follow.

\subsection{Planck}
\label{sec:Planck}
\begin{figure} [ht]
   \begin{center}
   \begin{tabular}{c} 
   \includegraphics[width=\textwidth]{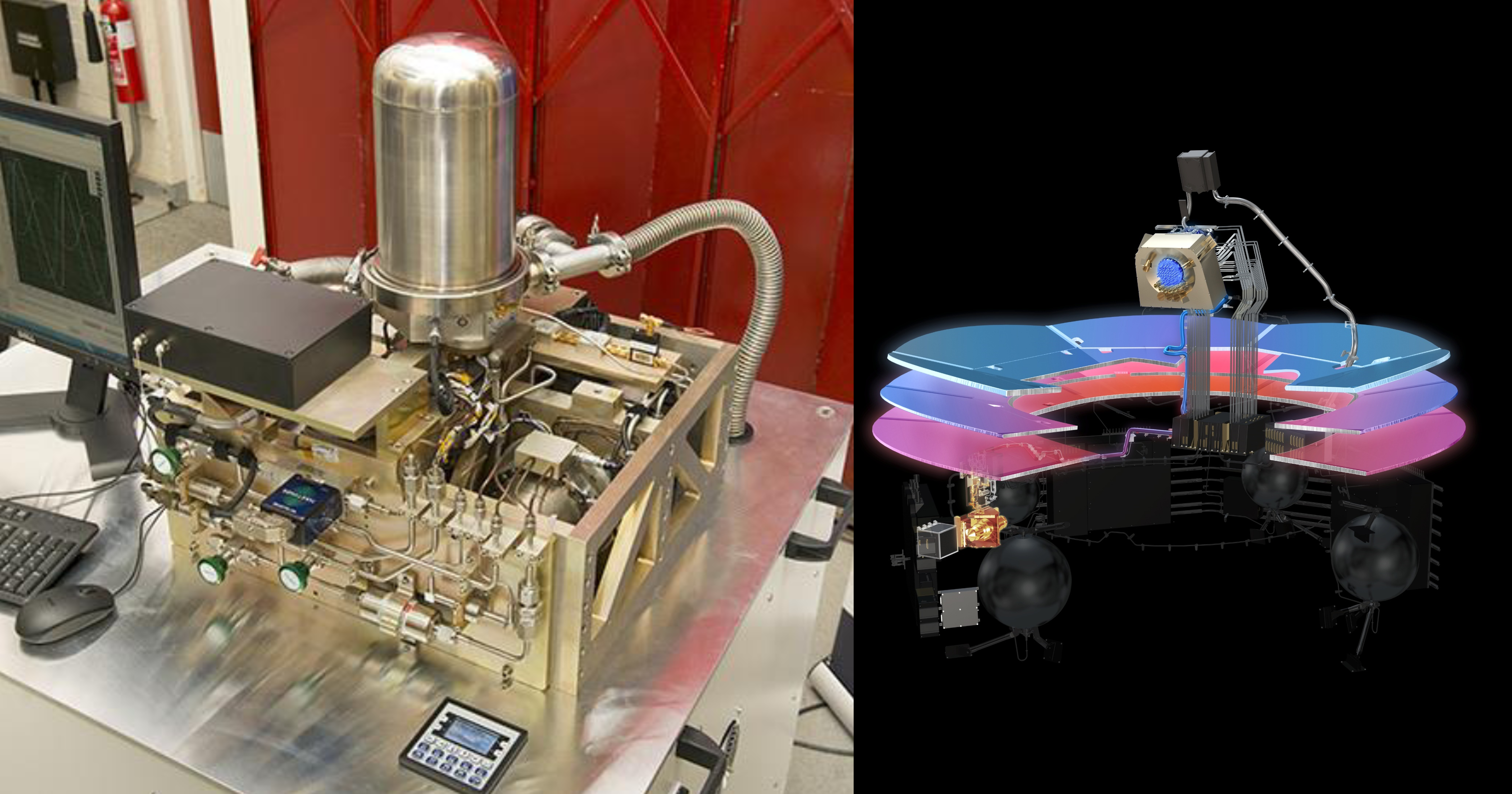}
	\end{tabular}
	\end{center}
   \caption[An identical model of the space qualified 4\,K JT cryocooler flown on Planck (left)\cite{hymatic-planck} and the active cooling system on board the Planck spacecraft (right)\cite{planck-website}.] 
   { \label{fig:planck} 
The space qualified 4\,K JT cryocooler flown on Planck (left)\cite{hymatic-planck} and the active cooling system on board the Planck spacecraft (right)\cite{planck-website}.}
   \end{figure}

Planck was a European Space Agency (ESA) space observatory that studied the cosmic microwave background at microwave and infrared frequencies and was launched in 2009 along with Herschel \cite{Morgante_2009}. The sorption cooler that was utilized for cryocooling had two primary functions: it cooled the Planck Low Frequency Instrument (LFI) to temperatures below 22.5\,K and served as an initial cooling stage for the 4\,K JT cooler used by the High Frequency Instrument (HFI). The required temperature stability was under 100\,mK for the LFI interface and below 450\,mK for the HFI. These coolers were engineered to provide 650\,mW for the LFI and 200\,mW for the HFI, with an overall input power of 470\,W, excluding electronic components. The effectiveness of these coolers largely depended on the temperatures at the compressor interface and the final pre-cooling stage \cite{planck-flight-testing}. The Rutherford Appleton Laboratory (RAL), supported by the European Space Agency (ESA), developed the 4\,K cryocooler used for the HFI 4\,K stage, employing helium-4 \cite{ral-4k_1997, ral-4k_1999}. Figure \ref{fig:planck} shows an identical cooler to the space qualified 4\,K JT cryocooler that was flown on Planck (left)\cite{hymatic-planck} and the active cooling system on board the Planck spacecraft (right)\cite{planck-website}, where the three V-grooves for passive cooling are color-highlighted, indicating progressively lower temperatures from bottom to top, and the closed-cycle 4\,K JT cooler integrated with the focal plane unit (FPU) of the HFI is shown. The compressors, highlighted in yellow in the bottom left, pump helium-4 gas to the FPU cold end and undergoes JT expansion producing small quantities of liquid helium that cool the HFI FPU \cite{planck-website, hymatic-quantum}.

\subsection{ACTDP and the James Webb Space Telescope}
\label{sec:JWST}
\subsubsection{Advanced Cooler Technology Development Program (ACTDP)}
In 2001, NASA initiated funding for the Advanced Cooler Technology Development Program (ACTDP) to develop long lifetime cryocoolers in the 4-20\,K temperature range, under the leadership of the NASA Jet Propulsion Laboratory and in collaboration with the NASA Goddard Space Flight Center \cite{ronross_2002, ron-ross-50years, mdp_2009, ronross_icc_2022}. The James Webb Space Telescope (JWST), known initially at the time as the Next Generation Space Telescope (NGST), was under conception and set to replace the Hubble Space Telescope (HST). Interest arose in the astrophysics community for a midwave infrared instrument with significantly longer wavelength capabilities than the HST, through the use of Si:As detectors cooled to 6\,K. At the time, this was a challenging cold temperature to cool to in space, particularly given that no 6\,K cryocooler designs with spaceflight maturity existed yet \cite{ronross_icc_2022}. 

\begin{figure} [ht]
   \begin{center}
   \begin{tabular}{c} 
   \includegraphics[height=8.5cm]{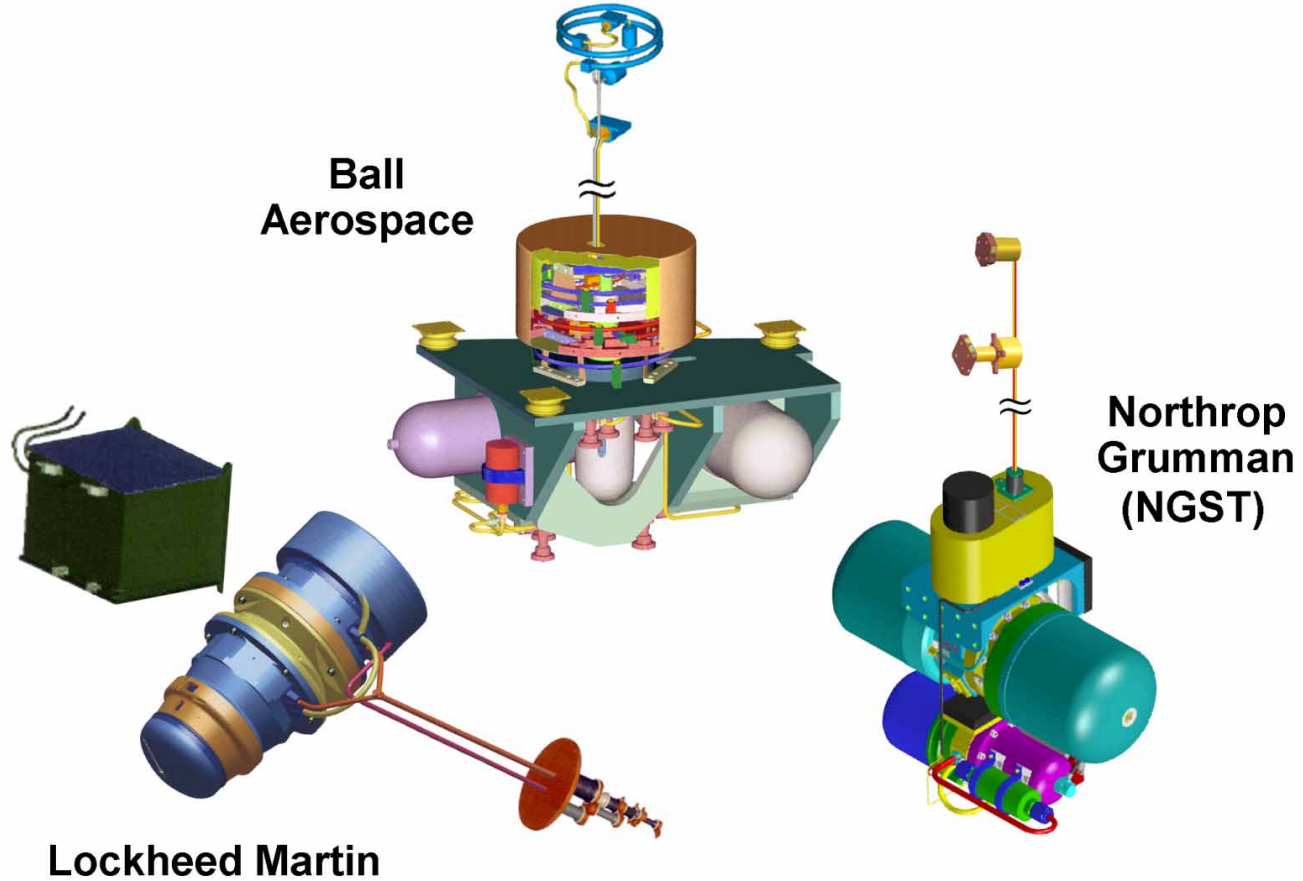}
	\end{tabular}
	\end{center}
   \caption[   6\,K cryocoolers developed by Ball Aerospace (now BAE Systems), Northrop Grumman, and Lockheed Martin \cite{ronross_icc_2022}.] 
   { \label{fig:ACTDPcoolers} 
ACTDP 6\,K cryocoolers developed by Ball Aerospace (now BAE Systems), Northrop Grumman, and Lockheed Martin \cite{ronross_icc_2022}.
}
   \end{figure}

Contracts in response to a request for 6\,K/18\,K cryocoolers were issued to meet these instrument cooling needs. This instrument eventually became the Mid-Infrared Instrument (MIRI). Four contracts were issued to TRW (now Northrop Grumman), Ball Aerospace (now BAE Systems), Lockheed Martin, and Creare \cite{ron-ross-50years}. The key objectives were to develop the aforementioned cooling stages of interest, operate remote cold heads, and minimize electromagnetic interference (EMI) and vibration noise. The top-level requirements included: a cooling power of 30\,mW at 6\,K and 150\,mW at the 18\,K cooling stage; less than 200\,W of input power; less than 40\,kg cryocooling system mass; 5-25 meter cold end deployment length accommodation; a lifetime performance exceeding 10 years; and low generated vibration and EMI \cite{ronross_icc_2022}. Figure \ref{fig:ACTDPcoolers} depicts the ACTDP 6\,K cryocooler designs by Ball Aerospace (now BAE Systems), Lockheed Martin, and Northrop Grumman. The Lockheed Martin cryocooler is a 4-stage pulse tube cryocooler; the other two are Stirling-JT cryo-chains, and were heavily baselined on the RAL Planck 4\,K cryocooler (see section \ref{sec:Planck}) \cite{ronross_icc_2022, hymatic-planck}. Additionally, Creare conducted initial studies on the development of a 4\,K Reverse Turbo-Brayton (RTB) cryocooler, however, this cooler was not taken forward for further design. This is due to the overwhelming challenge in miniaturizing RTBs for spaceflight \cite{MdP_workshop_2002}. Presently, Creare is continuing these developments for aerospace applications through NASA SBIR funding following later breakthroughs in miniaturization \cite{creare-4K-RTB} (see section \ref{sec:actdp-4k} for further discussion).

\subsubsection{JWST MIRI}
\begin{figure} [ht]
   \begin{center}
   \begin{tabular}{c} 
   \includegraphics[height=8.5cm]{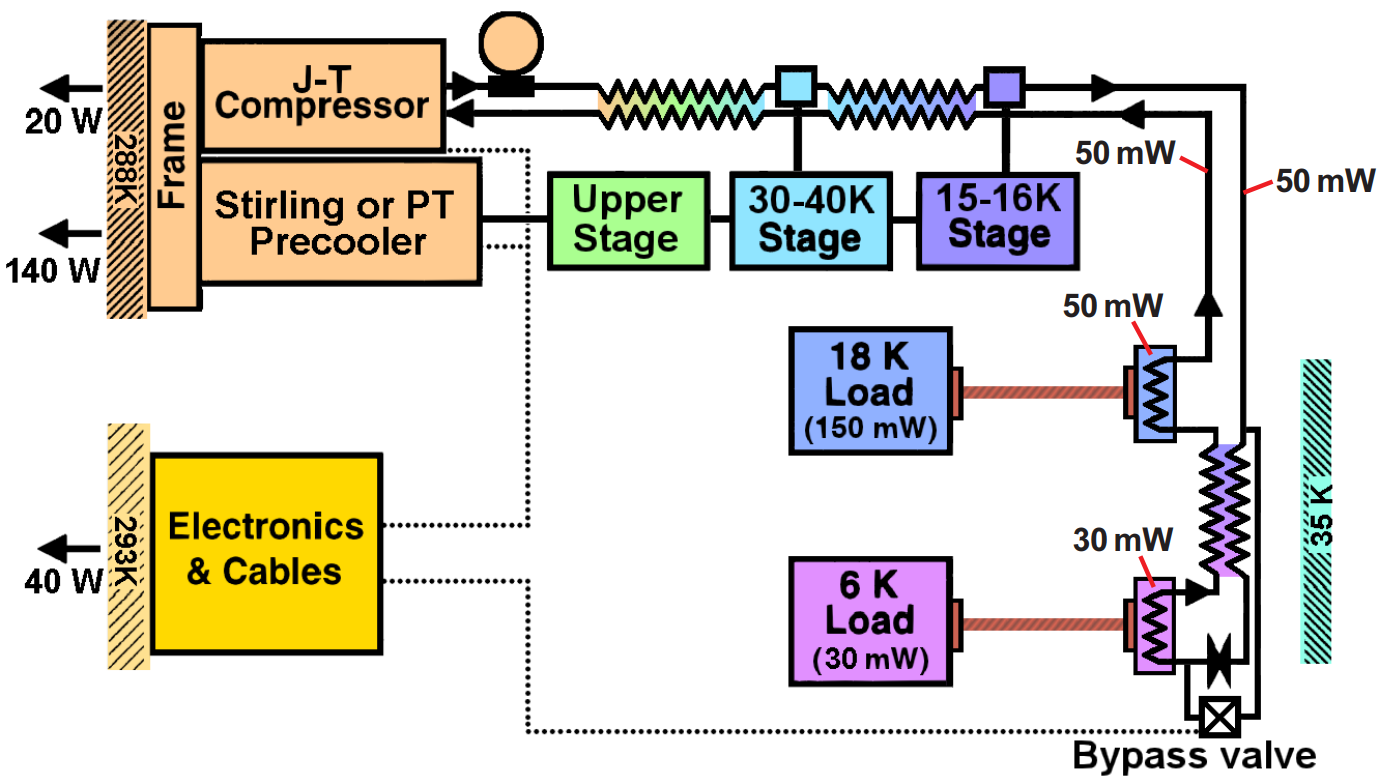}
	\end{tabular}
	\end{center}
   \caption[General architecture of Stirling/PT pre-cooling and JT cooling to achieve an 18\,K and a 6\,K cooling stage \cite{ron-ross-50years}.] 
   { \label{fig:MIRIflow} 
General architecture of Stirling/PT pre-cooling and JT cooling to achieve an 18\,K and a 6\,K cooling stage developed within the ACTDP for JWST MIRI\cite{ron-ross-50years}.
}
   \end{figure}

These developments ultimately led to the Northrop Grumman JWST MIRI cryocooler being selected, particularly due to the fact that the Stirling-JT design is favorable for remote cooling compared to the 4-stage pulse tube cryocooler \cite{ronross_icc_2022}. Figure \ref{fig:MIRIflow} shows the thermodynamic process for achieving cooling at the 18\,K and 6\,K stages. The TRL 7 JWST/MIRI cryocooler was engineered to cool a primary load to 6.2\,K and an intercept load to 18\,K. This was achieved using a pre-cooled helium JT loop, which compresses helium-4 from 4 to 12\,bar, pre-cools it to 18\,K with a three-stage pulse tube cryocooler, and then circulates it to a remote heat intercept at 18\,K and an isenthalpic expander for 6.2\,K cooling \cite{ronross_icc_2022}. 

\subsubsection{Towards a MIRI-Based 4\,K Space Cooler}
\label{sec:actdp-4k}

Presently, US cryocoolers capable of reaching 4.5\,K are generally considered to be at TRL 4-5, having been demonstrated in a laboratory setting under the ACTDP \cite{ross_2006}. For further development of the JWST MIRI cryocooler architecture to reach a cold temperature of 4\,K, the return pressure requires decreasing from 4\,bar to 1\,bar. To sustain the necessary mass flow, the compressor's swept volume needs to be increased to account for the lower density of helium at reduced pressure, and the pressure ratio must be raised to maintain the pressure drop. This can be achieved by enhancing the JT compressor used in the MIRI cooler, potentially by adding a second JT compressor with larger pistons to serve as the first stage in a two-stage compressor setup \cite{origins-tech-dev, mdp_origins}. The moving components of a 4.5\,K cooler, such as expanders, are similar or identical to those that have flown before. Further development aims to maximize cooling power per input power for small cooling loads (50-200 mW) while reducing mass and minimizing exported vibration \cite{ronross_icc_2022, mdp_2009}. 

Presently, the BAE Systems (formerly Ball Aerospace) 4.5\,K cooler design employs a similar architecture to the JWST/MIRI cryocooler, but with a Stirling-type displacer instead of a pulse tube and a JT with a long loop for remote cooling. This design requires a system-level demonstration to reach TRL 5 \cite{glaister_2007}. Lockheed Martin's four-stage pulse tube cryocooler has demonstrated the required heat removal at the necessary temperatures for SPICE (formerly the Origins Telescope) in a TRL 4 design \cite{olson_2005, origins-tech-dev}. 

Creare is currently funded through an SBIR Phase II contract to demonstrate a 4.5\,K turbo-expander, which is the lowest TRL component of the Creare design, approaching TRL 4. Creare is also developing a 4.5\,K stage for its miniature RTB cryocoolers. This expansion stage is similar to one used for a single-stage cooler flown on the Near Infrared Camera and Multi-Object Spectrometer (NICMOS) on the HST and the two-stage 10\,K design currently at TRL 5. The miniature RTB cryocoolers being developed at Creare indicate high efficiency and reliability with reduced exported vibration \cite{creare-4K-RTB}. These coolers are at TRL 6 for 80\,K, TRL 4-5 for 10\,K, and TRL 3 for 4 K operation \cite{origins-tech-dev}.

\subsection{4\,K Pulse Tube Cryocoolers}
\label{sec:4KPT}
\begin{figure} [ht]
   \begin{center}
   \begin{tabular}{c} 
   \includegraphics[height=7.5cm]{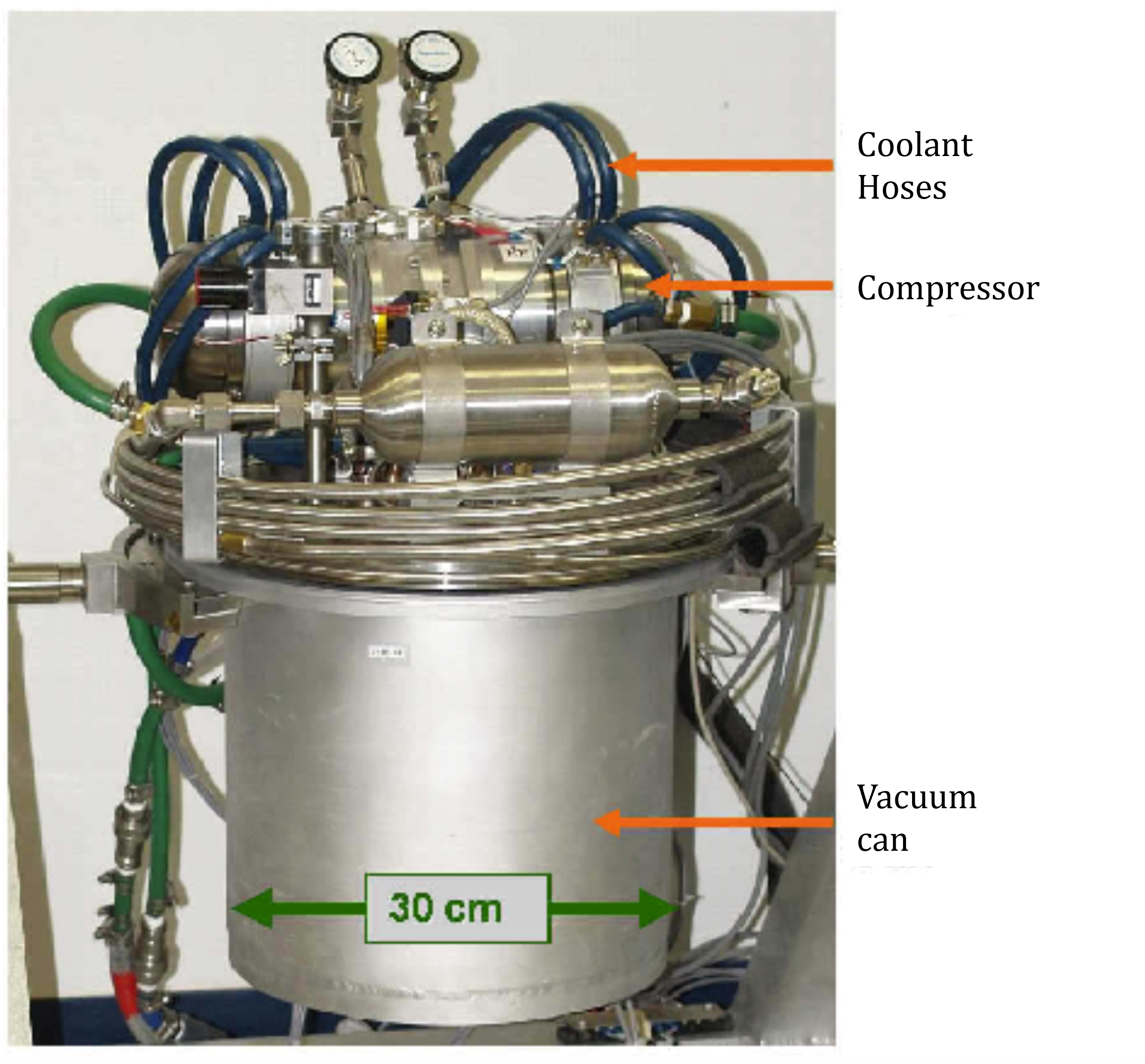}
	\end{tabular}
	\end{center}
   \caption[The complete cryocooler and cryostat laboratory testing setup for the HYPRES cooler using Lockheed Martin's 4\,K 4-stage pulse tube cryocooler \cite{LM_4K_PT2_HYPRES}.] 
   { \label{fig:hyprescooler} 
The complete relatively compact cryocooler and cryostat laboratory testing setup for the HYPRES cooler using Lockheed Martin's 4\,K 4-stage pulse tube cryocooler \cite{LM_4K_PT2_HYPRES}.
}
   \end{figure}

As addressed in section \ref{sec:actdp-4k}, efforts have been made to procure a 4\,K spaceflight pulse tube cryocooler, however, such a cryocooler architecture has not yet flown in space to date. Pulse tube cryocoolers run regenerative cycles, are driven by either Stirling or Gifford-McMahon compressors, and are generally considered favorable from a vibration perspective, given that they employ orifices or inertance tubes at the warm end for phase control as opposed to actively-driven displacers that import vibrations into the cold end where the detector is integrated \cite{passive1, passive2, rana2020}. Pulse tube cryocoolers for space applications typically operate at high frequencies for optimized refrigeration, however, high frequencies limit the lowest temperature achievable due to regenerator performance degradation. Research has been conducted to develop a 4\,K pulse tube cold finger that couples with traditional pre-cooling stages or multi-stage pulse tube cryocoolers that reach an eventual 4\,K cold stage \cite{japan_4K_PT, CEA_4K_PT, LM_4K_PT1}. In conventional multistage configurations, the cryocooler needs to operate at 1-2\,Hz in order to achieve temperatures around 4\,K. By utilizing a traditional spaceflight cryocooler at high-temperature stages and operating it at a high frequency, whilst thermally coupled to a 4\,K stage run at low frequency, a conceptual compact spaceflight design has been developed that can achieve 4\,K cooling with a reasonable efficiency and cooling power \cite{japan_4K_PT}. Such a cooler remains below TRL 4, however, based on reported literature. Lockheed Martin first achieved a cold temperature of 4.2\,K with a 4-stage pulse tube cryocooler configuration using helium-3 as the working fluid, leading to the development of their HYPRES 4\,K pulse tube cryocooler for superconducting electronics \cite{LM_4K_PT1, Cao_4K_PT_2011, NIST_4K_PT, LM_4K_PT2_HYPRES}.This was borne out of previous series of efforts by Lockheed Martin in the early 2000's to reach 10\,K with similar pulse tube cooler architecture \cite{LM_10K_PT}. Moreover, multi-stage pulse tube coolers that reach 4\,K using helium-4, instead of the much more rare helium-3, have since been demonstrated \cite{Zhi_4K_PT_He4_2013}. Figure \ref{fig:hyprescooler} shows the HYPRES complete cryocooler and cryostat laboratory setup \cite{LM_4K_PT2_HYPRES}. The Lockheed Martin ACTDP cryocooler development continues\cite{olson_2005, origins-tech-dev}, as discussed in section \ref{sec:actdp-4k}. There remains a need to reach higher TRLs of these architectures for space applications. CEA Grenoble and CNES have made significant progress in developing a 4\,K pulse tube cryocooler within the framework of the ATHENA mission, which is discussed further in section \ref{sec:athena}.

\subsection{JAXA/SHI 4\,K Closed Cycle Space Coolers}
\label{sec:SHI}
Japan has been at the forefront of developing 4\,K closed-cycle cryocoolers for space missions, being responsible for the three fully closed-cycle cryocooler systems operating at 4-4.5\,K to have ever flown in space. These were in space missions led by JAXA where the cryocoolers were produced by Sumitomo Heavy Industries (SHI) \cite{narasaki2012}.

\begin{figure} [ht]
   \begin{center}
   \begin{tabular}{c} 
   \includegraphics[width=\textwidth]{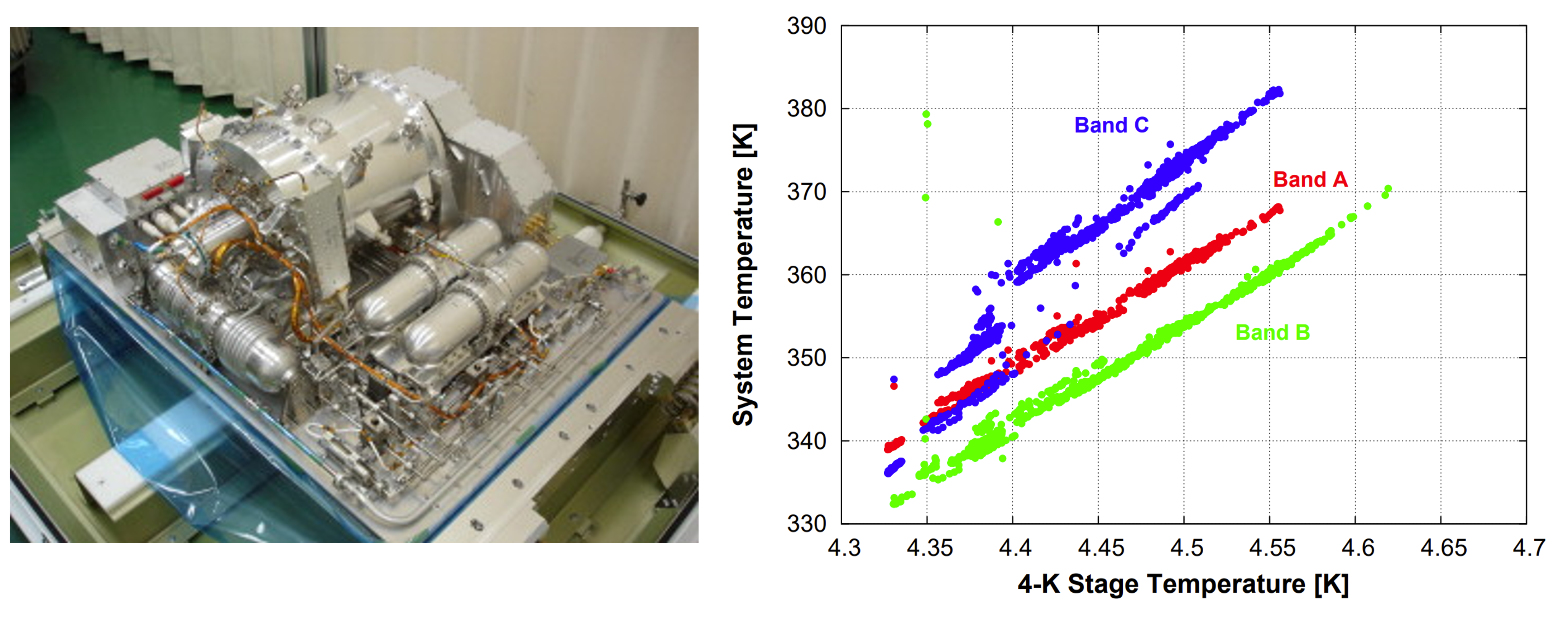}
	\end{tabular}
	\end{center}
   \caption[The cryocooling system that flew on-board JEM/SMILES (left) \cite{otsuka2010} and the relationship between the system noise temperature and the temperature of the 4\,K stage, in-orbit (right) \cite{ochiai2010}] 
   { \label{fig:jem-smiles} 
The cryocooling system that flew on-board JEM/SMILES (left) \cite{otsuka2010} and the relationship between the system noise temperature and the temperature of the 4\,K stage, in-orbit (right) \cite{ochiai2010}.
}
   \end{figure}

The Superconducting Submillimeter Wave Limb Emission Sounder (SMILES) mission was the first space demonstration of the JAXA/SHI 4\,K fully closed-cycle cryocooling system, on the International Space Station (ISS) \cite{masuko1997}. The purpose of the mission was to study ozone depletion-related chemical processes, in collaboration with the National Institute of Information and Communications Technology (NICT) and JAXA. The cryocooler consisted of the (now signature) SHI architecture comprised of a two-stage Stirling cryocooler and a JT cooler and cooled 650,GHz Superconductor Insulator Superconductor (SIS) mixers and low-noise High Electron Mobility Transistor (HEMT) amplifiers to 4.5\,K \cite{narasaki2012, ochiai2010}. The JT lower stage cooler produced 20\,mW of cooling at 4.5\,K using helium-4 and the 2-stage Stirling pre-cooler cooled the HEMT amplifiers at 20\,K and 100\,K stage \cite{inatani2005, otsuka2010, yoshida2016}. The total input power required for the cooler was approximately 140\,W \cite{mdp_2009}. Figure \ref{fig:jem-smiles} shows the cryocooling system that flew on-board JEM/SMILES (left) \cite{otsuka2010} and the relationship between the system noise temperature and the temperature of the 4\,K stage, in-orbit (right) \cite{ochiai2010}. The Band A and Band B system noise temperatures are well correlated with the 4\,K stage, however, not aligned for Band C. Operational parameters have been changed in-flight that have caused the latter; a heater switch-off led to increased gain increment in one of the amplifiers leading to the increase in system noise temperature. Additionally, the Stirling compressor operation frequency and voltage input was altered which can lead to affects in system noise temperature also \cite{ochiai2010}.

\begin{figure}[ht]
    \centering
    \makebox[\textwidth][c]{
        \includegraphics[width=\textwidth]{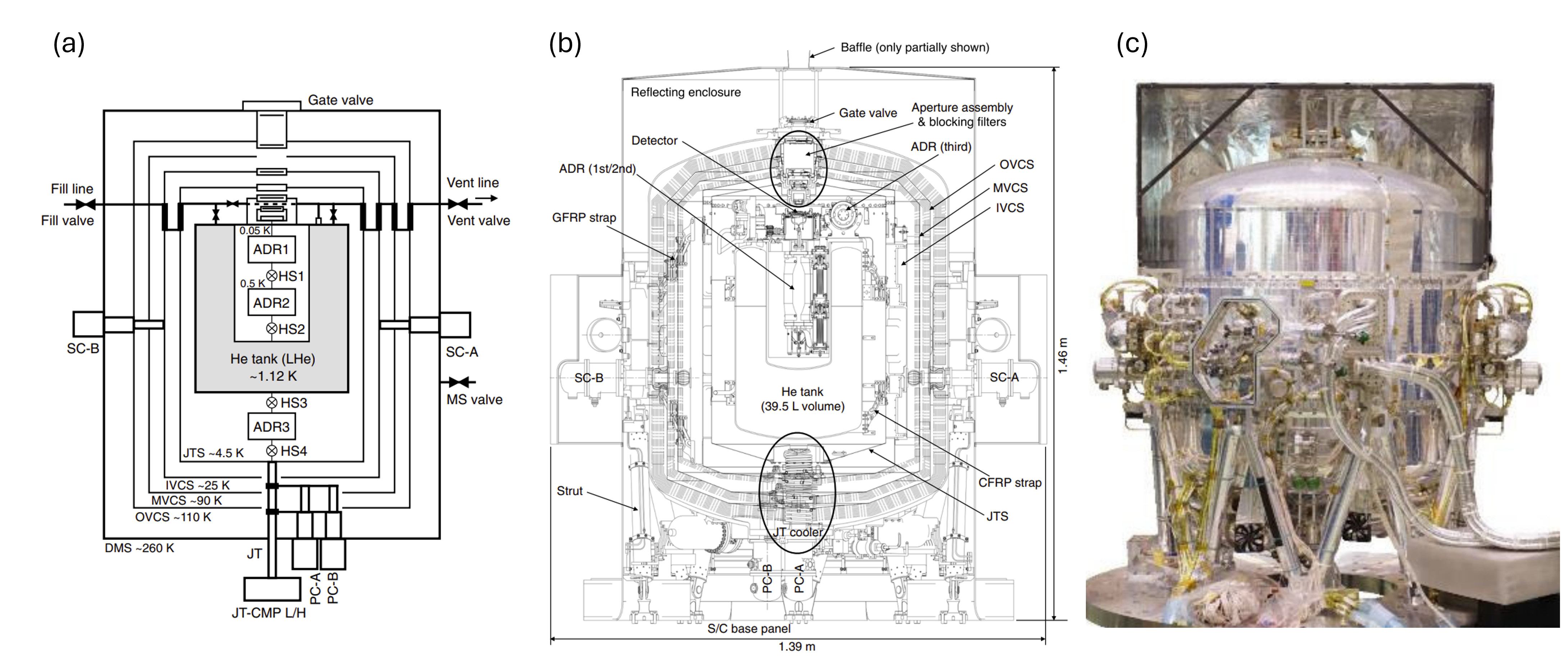}
    }
    \caption{The JAXA/SHI Hitomi cryocooling system is presented, showing (a) the conceptual design of the SXS cryocooling \cite{fujimoto2018}, (b) the cross-sectional view of the SXS dewar\cite{fujimoto2018}, and (c) the physical image of the full Hitomi cryocooling system \cite{yoshida2018}.}
    \label{fig:hitomi-abc}
\end{figure}

Following the success of SMILES, the Hitomi (ASTRO-H) mission, a JAXA-NASA X-ray astronomy mission, employed a similar system with the same architecture for the 4\,K and warmer stages, developed by SHI for its Soft X-ray Spectrometer (SXS). The lowest cooling stage required temperatures below 1.3\,K. The cryocooler has a lifetime goal of 5 years' operation \cite{origins-tech-dev}. Although Hitomi, launched in February 2016, only operated for a month due to control issues and loss of communication, its instruments, including the SXS and its cryocooler, functioned as intended \cite{takahashi2018, yoshida2018, fujimoto2018}. Figure \ref{fig:hitomi-abc} presents the Hitomi cryocooling system, showing (a) the conceptual design of the SXS cryocooling \cite{fujimoto2018}, (b) the cross-sectional view of the SXS dewar\cite{fujimoto2018}, and (c) the physical image of the Hitomi cryocooling system \cite{yoshida2018}.

Thereafter, JAXA/SHI began developments for the 20\,K, 4.5\,K and 1.7\,K coolers needed for the ESA Space Infrared Telescope for Cosmology and Astrophysics (SPICA) space telescope mission. The SHI 4.5\,K cooler was enhanced to increased cooling power from 20\,mW to 50\,mW with an input power of 145\,W \cite{mdp_2009, sugita2006}. These enhancements were achieved due to advancements made in the 20\,K pre-cooler performance, whereby the cooler was able to produce an increased 325\,mW of cooling power at 20\,K. Utilizing helium-3 in the JT cooler allowed for 16\,mW of cooling power at the 1.7\,K to be achieved \cite{sugita2008, nakagawa2020}. Unfortunately, the SPICA mission was eventually not selected by ESA for a Mid-Class mission as part of its Cosmic Vision Programme (M5), due to financial constraints arising from the increasing cost of the overall mission \cite{esa_spica}. JAXA's `Lite (Light) satellite for the studies of B-mode polarization and Inflation from cosmic background Radiation Detection (LiteBIRD)' mission will aim to leverage the developments in cryogenic architecture for SPICA using the SHI 4\,K Stirling-JT cooler \cite{hazumi2021}.

The X-Ray Imaging and Spectroscopy Mission (XRISM), launched in September 2023 as a recovery mission for Hitomi, carries on-board the Resolve instrument, consisting of a microcalorimeter and a similar SHI-manufactured cryocooling system to that flown on Hitomi. The cryocooling system was kept as similar to Hitomi as possible, given its successful qualification and performance, in order to reduce time taken in resuming the scientific aims that Hitomi had sought to pursue \cite{ishisaki2022, ezoe2020}. The XRISM cryocooler is currently in-flight and further details on on-board performance are expected to be disseminated in the astronomy community in due course \cite{mamura2023}.

The SHI 4.5\,K cryocooler flown on Hitomi and XRISM had a lifetime goal of 5 years. Lifetime is expected to be limited by bearing friction and contamination. Reliability and lifetime can sought to be improved by using a non-contacting suspension system (similar to displacers) and by improving the cleanliness of the critical internal parts and working fluid. This will extend expected life from 5 to 10 or more years. This development is in progress at SHI \cite{origins-tech-dev}. 

\subsection{ATHENA}
\label{sec:athena}
The Advanced Telescope for High-Energy Astrophysics (ATHENA) is an X-ray space observatory planned for launch in the mid 2030's and will be the second Large Class mission in the ESA Cosmic Vision program. ATHENA plans to map hot gas structures and identify and study supermassive black holes amongst other aims \cite{barcons2015}. From the cryocooling point of view, the X-ray Integral Field Unit (X-IFU) instrument on ATHENA, which utilizes Transition Edge Sensor (TES) detectors to be cooled to 50\,mK, will be one of the most challenging cryocooled instruments to ever fly in space \cite{tirolienICC}. The full cryo-chain is currently planned to include a 15\,K JT precooling stage, a 4\,K stage, a 2\,K stage, and the sub-Kelvin cooler to 50\,mK \cite{CEA_4K_PT, tirolienICC, barret2018, duval2012, crook2016}. NASA is now responsible for the 4\,K cryocooler, with US-based cryocooler and V-groove passive cooling combinations currently being considered for the 4\,K cryo-chain. 

\section{BHEX Cryocooling System}
\subsection{System Layout}
\begin{figure}[ht]
    \centering
    \makebox[\textwidth][c]{
        \includegraphics[width=0.5\textwidth]{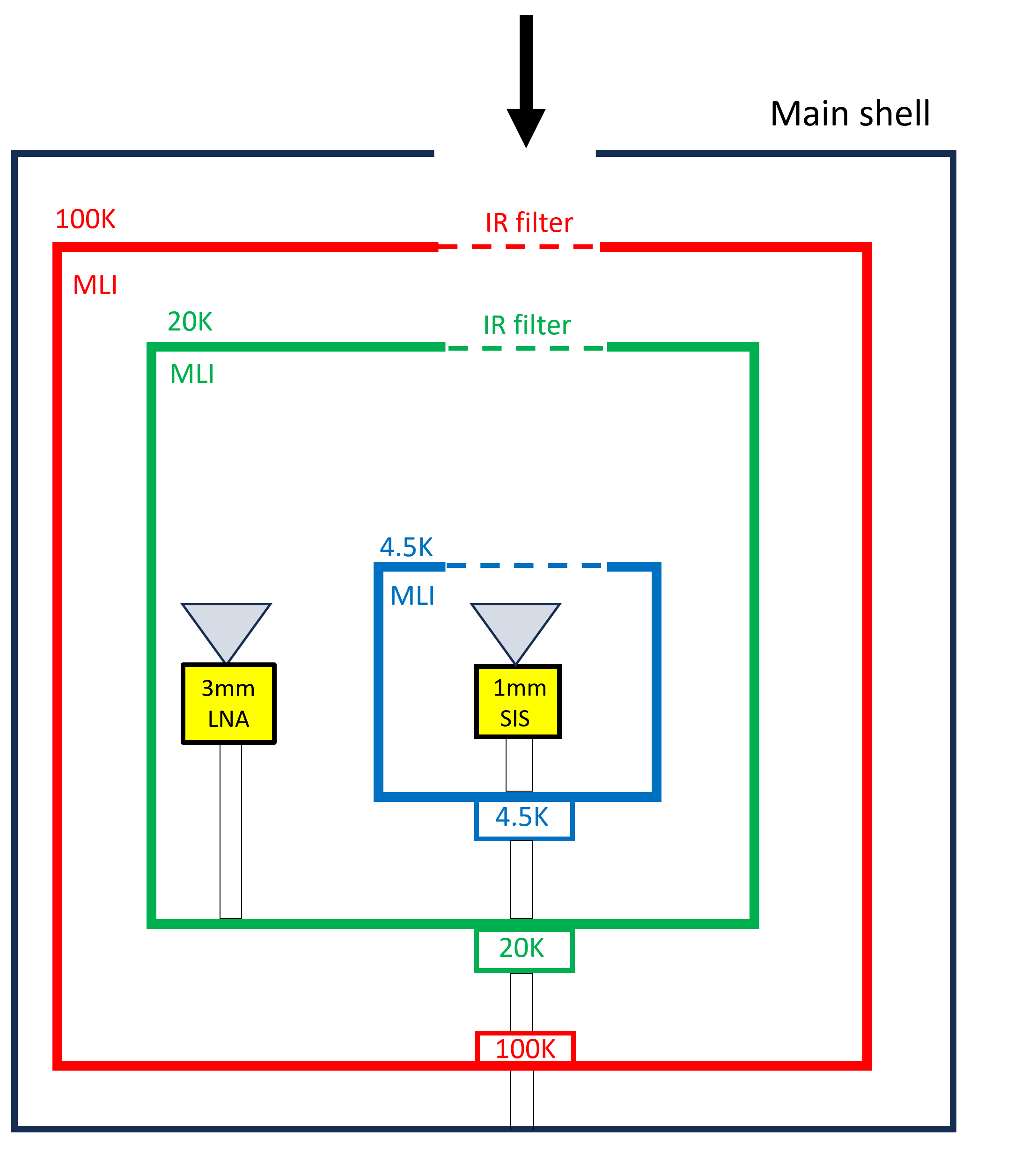}
    }
    \caption{The BHEX cryocooling stages integrated with the receiver system.}
    \label{fig:bhex-layout-nocryo}
\end{figure}

The layout for the BHEX receiver system \cite{BHEX_Tong_2024} and the cooling stages required are shown in Figure \ref{fig:bhex-layout-nocryo}. The SIS receiver, referred to as the High Band Receiver,\cite{BHEX_Tong_2024, BHEX_Marrone_2024} is housed within the 4.5\,K stage and the Low Band Receiver is housed within the 20\,K. An intermediate stage of 100\,K is achieved through pre-cooling and the potential of passively cooling to warmer stages is currently being explored. Multi-Layer Insulation (MLI) is utilized to shroud the stage layers and provide thermal insulation between each stage. Liquid helium tanks, as were used in all previous closed-cycle 4\,K cryocoolers to have flown in space, by JAXA/SHI, will not be used in the BHEX mission due to their added mass and volume, and the fact that the cooling requirements as outlined in section \ref{sec:reqs} do not necessitate them. 

In order to permit cooling from the cryogenic system warm end, a thermally linked radiator between the cryocooler and the spacecraft would need to be facing deep space at all times for passive radiative heat rejection to 3\,K space. Moreover, a second and smaller radiator, which could be Earth-facing, may be required. The cryocooler electronics drive box typically cannot exceed a maximum temperature of approximately 40$^o$C, where usually their temperature is maintained both during on-ground testing and in-flight at around 30$^o$C. The cryocooler warm end compressors would also need to be maintained at approximately the same temperature. These elements will require dedicated thermal control design.

The BHEX cryocooling system is currently cryo-chain architecture agnostic - meaning, the BHEX team is currently in the process of selecting a cryocooler vendor and baselining the design accordingly \cite{CSA2024}.

\subsection{Spaceflight Testing and Qualification}
The cryocooling unit will be a protoflight unit - one single unit that is taken through each stage of the space mission life cycle, including design, performance testing and optimization, qualification testing, and in-flight use. No redundant cryocoolers will be present. Thankfully, the majority of components (Stirling cryocoolers, pulse tube cryocoolers, JT cryocoolers, thermal linking hardware, drive electronics etc.) that make up the various technologies that achieve 4\,K cryocooling are technologically mature, where additionally, the full JAXA/SHI 4\,K cryocooling systems have successfully flown multiple times (see section \ref{sec:SHI}). This is beneficial in the case of a SMEX mission where the timeline is comparatively short ($\sim$5 years).

The testing to be undergone by the cryocooling system begins at the component level, where the individual cryocoolers within the chosen cryo-chain are performance tested. Key parameters such as their cooling power, efficiency, and cold end temperature are verified prior to integration into the full cryo-chain. Thereafter, the cryo-chain is integrated into the vessel and assembled with dedicated performance testing metrology. Given that in order to comply with the mass budget of the SMEX payload, a large cryostat will not be being utilized, a dedicated cryo-chamber vessel for on-ground testing purposes will be required. Most cryocooler vendors will have these facilities available for testing at this level. The cryocooling system undergoes functional component testing as a full unit. This is followed by vibration and thermal vacuum chamber (TVAC) testing, including cycling tests to ensure contamination procedures are effective, before acceptance review takes place and the subsystem is transported to the spacecraft vendor for integration with the receiver system, spacecraft and instrument as a whole. The aforementioned serves as an approximate and brief overview. Spaceflight qualification standards are to be strictly adhered to at each stage of the rigorous testing processes \cite{GSFC2023}. 

\subsection{Design Trades}
Table \ref{tab:trades} gives examples of some of the key trade-offs to be made for the BHEX cryocooling system. Several design trade-offs are essential for both the instrument operation and the cryogenic system to ensure the scientific objectives are achieved while scaling the cryocooling system to fit within the constraints of a SMEX mission. For instance, the SIS mixers must be operated at a temperature no higher than 4.5\,K to achieve the necessary sensitivity for imaging the photon ring; this requirement determines the power input into the cryocooler and the cooling power needed from the cryocooler. Additionally, continuous operation of the cryocooler is required to maintain receiver stability, as opposed to intermittent operation, which prevents power-saving measures. However, this approach also reduces the risk of contamination (a primary cause of cryocooler failure) and eliminates the need for cool-down and temperature stabilization before each observation cycle. Designing the cryogenic system for BHEX involves balancing mass, power, and cost reduction to meet SMEX constraints while ensuring optimal cooling performance and minimizing failure risks.

\begin{table}[ht]
\caption{Examples of BHEX Cryocooling System Trades} 
\label{tab:trades}
\begin{center}       
\begin{tabular}{|p{0.25\textwidth}|p{0.3\textwidth}|p{0.35\textwidth}|} 
\hline
\rowcolor{blue!40}  \textbf{Design Choice} & \textbf{Rationale} & \textbf{Penalty} \\
\hline
\rowcolor{blue!5} 4.5\,K SIS cold temperature & Enhanced sensitivity & Increased input power into cryogenic system \\
\hline
\rowcolor{blue!5} Continuous operation of cryocooler instead of on/off operation & Reduction of contamination risk and cooldown time after switching on during observation & Larger input power into cryogenic system and power drawn within same timeframe as all other subsystems \\
\hline
\rowcolor{blue!5} No large vessel cryostat & Reduction of overall mass and increased simplicity of system integration & Dedicated testing vessel required for ground testing which is design and cost intensive \\
\hline
\end{tabular}
\end{center}
\end{table}

\section{Summary}
The BHEX mission sets out to capture the first ever image of the black hole photon ring, study key properties of black holes, and experimentally validate predictions of General Relativity amongst many other exciting scientific aims. This will be made possible thanks to dedicated instrumentation and spaceflight hardware, including cryogenic receivers offering quantum-limited sensitivities across a wide frequency coverage. The BHEX cryocooling system will cool the receiver system to 20\,K and 4.5\,K stages. The last decades have seen monumental developments in 4-10\,K spaceflight cryocoolers, through the European Planck mission, the ACTDP and JWST MIRI endeavors, and the range of Japanese missions flying closed-cycle 4\,K stage cryocoolers. The BHEX mission will leverage these established and ever-improving space instrumentation capabilities.

\appendix

\acknowledgments 

Technical and concept studies for BHEX have been supported by the Smithsonian Astrophysical Observatory,
by the Internal Research and Development (IRAD) program at NASA Goddard Space Flight Center, by the
University of Arizona Space Institute, and by the ULVAC-Hayashi Seed Fund from the MIT-Japan Program
at MIT International Science and Technology Initiatives (MISTI). We acknowledge financial support from the
Brinson Foundation, the Gordon and Betty Moore Foundation (GBMF-10423), and the National Science Foundation (AST-2307887, AST-2107681, AST-1935980, and AST-2034306). This work was supported by the Black Hole Initiative at Harvard University, which is funded by grants from the John Templeton Foundation and the
Gordon and Betty Moore Foundation to Harvard University. BHEX is funded in part by generous support from Mr. Michael Tuteur and Amy Tuteur, MD. BHEX is supported by initial funding from Fred Ehrsam. HR is supported by the Schmidt Science Fellows.

\bibliography{report.bib} 
\bibliographystyle{spiebib} 

\end{document}